\documentclass[prd,twocolumn,nofootinbib]{revtex4}
\usepackage{bm}
\usepackage{amssymb}
\usepackage{graphicx}
\usepackage{epstopdf}
\usepackage{mathrsfs}
\usepackage{amsmath}

\usepackage[utf8]{inputenc}
\usepackage[english]{babel}
\usepackage{xcolor}

\begin{document}
\title{Repeated Ringing of the Black Hole's Bell: Quasi-Normal Bursts from Highly Eccentric, Extreme Mass-Ratio Binaries} 
\author{Nur E.M.~Rifat$^1$, Gaurav Khanna$^1$, and Lior M.~Burko$^2$}
\affiliation{
$^1$ Department of Physics, University of Massachusetts, Dartmouth, Massachusetts  02747, USA\\
$^2$ Theiss Research, La Jolla, California 92037, USA}
\date{October 8, 2019}
\begin{abstract}
Recent studies of scalar and gravitational waveforms from high-eccentricity, extreme mass-ratio black hole binaries show the presence of quasi-normal bursts (QNB) -- lingering high frequency decaying oscillations (also known as ``wiggles'') -- soon after each periapsis passage. One puzzle associated with these QNB    
is that in the case of a nearly-extreme rotating central black hole the frequency of the QNB has been found to be in a range which is lower than the corresponding range of relevant quasi-normal modes. We reproduce these results using a different approach and perform a detailed analysis to find evidence for the resolution of the puzzle and for the origin of the QNB. We find that the QNB frequency as measured at future null infinity evolves in (retarded) time and approaches the dominant quasi-normal frequency exponentially in time. We also show that the QNB amplitude decays inversely in (retarded) time.  We discuss the time dependence of both the QNB waveform frequency and its amplitude and argue that this behavior arises as a result of the excitation of many quasi-normal overtones and the summation thereof. 
\end{abstract}

\maketitle

\section{Introduction}


When the periapsis of the orbit of a compact object around a massive black hole gets close to the light ring of the latter, the compact object can excite the massive black hole's quasi-normal modes (QNM), which results in repeated bursts of high-frequency gravitational waves -- also known as ``wiggles" -- with each periapsis passage~\cite{thornburg,nasipak}. These repeated high-frequency bursts were first discovered in the energy flux at infinity for high-eccentricity, ``zoom-whirl'' orbits around a spinning black hole \cite{burko:2007}, and then rediscovered in the context of the horizon's shear response for perturbations of the massive black hole \cite{hughes:2016} and rediscovered again in the high-frequency oscillations in the self force on the compact object \cite{thornburg:2017}. 

The emitted radiation from a binary black hole systems in the extreme mass-ratio limit (EMRI) is  expected to be a source of gravitational radiation detectable by the space based Laser Interferometer Space Antenna (LISA). Much like ground-based detectors, accurate waveform signal templates will be needed for LISA data analysis for matched-filtering purposes and for precision parameter estimation. Thus, the development of computational models for EMRIs is an important research area of gravitational physics. 

In recent computations of scalar and gravitational waveforms from highly eccentric EMRIs repeated high frequency quasi-normal bursts (QNB) appear soon after the compact object passes through the orbital periapsis of the massive black hole~\cite{thornburg,nasipak}. QNMs are excited because on a high-eccentricity orbit, the compact object is able to get close to the light ring, 
and the high spin of the central black hole allows these modes to be long-lived. While excitations by a particle of QNM had been observed before (first seen for a marginally-bound particle which is scattered or absorbed by a Kerr black hole~\cite{kojima}), an unexpected 
feature emerges when the central black hole spins at a nearly-extreme rate: The frequency of the QNB waveform lies in a range that is lower than the relevant QNM frequencies~\cite{thornburg}. This curious phenomenon has been observed also in the self force on the compact object \cite{thornburg:2017} and in the shear response of the central black hole's horizon \cite{hughes:2016}. 
The origin of this feature has not been studied or discussed in depth, perhaps because it 
is notoriously difficult to numerically model nearly-extreme Kerr black holes. However, it was noted in Ref.~\cite{thornburg} that the effect may be due to ``some complicated collective behavior of the QNMs or some new physical effect."

Here, we use time-domain Kerr black hole perturbation theory -- i.e. we solve the Teukolsky equation in the time domain with a point-particle source -- to compute the gravitational waveform for a suitable high-eccentricity EMRI system and observe the generation of the QNB after the 
particle passes through the orbital periapsis. We perform a detailed analysis of the QNB waveform and find that the QNB frequencies are time dependent. Specifically, we obtain the precise time dependence of both its frequency 
and amplitude. Next, we show that precisely the same effect is also present in a much simpler, source-free scalar wave evolution in near-extremal Kerr 
spacetime background. This result suggests that the explanation for the time dependence of the frequency of the QNB  waveform is purely due to the quasi-normal spectrum 
of a nearly-extreme Kerr black hole and doesn't have much to do with the motion of a point-particle on a highly eccentric orbit. Next, we demonstrate the difference in the behavior of nearly-extreme and sub-extreme black holes, and specifically point out how high the spin of the black hole needs to be to observe the QNB phenomenon. This value of the spin is similar to that found in other aspects of the transition from sub-extreme to nearly-extreme black holes. We then suggest that the full explanation of the effect we consider in this work lies in the behavior of the QNM overtones of a near-extremal Kerr black hole.

\section{Point-particle black hole perturbation theory (ppBHPT)}


In the large mass-ratio limit of a black hole binary system, a study of the system's dynamics can be addressed using black hole perturbation theory. In that approach, the smaller black hole is modeled as a point particle with no internal structure, moving in the spacetime of the larger black hole. Gravitational radiation is computed by evolving the perturbations generated by this moving particle, by solving the Teukolsky master equation with particle-source~\cite{emri_code}. 
\begin{eqnarray}
\label{teuk0}
&&
-\left[\frac{(r^2 + a^2)^2 }{\Delta}-a^2\sin^2\theta\right]
        \partial_{tt}\Psi
-\frac{4 M a r}{\Delta}
        \partial_{t\phi}\Psi \nonumber \\
&&- 2s\left[r-\frac{M(r^2-a^2)}{\Delta}+ia\cos\theta\right]
        \partial_t\Psi\nonumber\\  
&&
+\,\Delta^{-s}\partial_r\left(\Delta^{s+1}\partial_r\Psi\right)
+\frac{1}{\sin\theta}\partial_\theta
\left(\sin\theta\partial_\theta\Psi\right)+\nonumber\\
&& \left[\frac{1}{\sin^2\theta}-\frac{a^2}{\Delta}\right] 
\partial_{\phi\phi}\Psi +\, 2s \left[\frac{a (r-M)}{\Delta} 
+ \frac{i \cos\theta}{\sin^2\theta}\right] \partial_\phi\Psi  \nonumber\\
&&- \left(s^2 \cot^2\theta - s \right) \Psi = -4\pi\left(r^2+a^2\cos^2\theta\right)T   ,
\end{eqnarray}
where 
$M$ is the mass of the black hole, $a$ its angular momentum per unit mass, 
$\Delta = r^2 - 2 M r + a^2$ and $s$ is the ``spin weight'' 
of the field. The $s=0$ case refers to a scalar field $\psi$, while the $s = -2$ version of this 
equation describes the radiative degrees of freedom of the gravitational field $\psi_4$ in the radiation 
zone, and is directly related to the Weyl curvature scalar as $\Psi = (r - ia\cos\theta)^4\psi_4$. 

Computing the Weyl curvature scalar $\psi_4$ from the motion of a small object around a Kerr 
black hole involves a two-step process. First, we compute the trajectory taken by the point-particle, 
and then we use that trajectory to compute the gravitational wave emission. For the first step, the 
particle's motion can simply be chosen to lie on a highly eccentric geodesic. In particular, we use 
the parameters $(a/M, p/M, e) = (0.999\,99, 2.918, 0.807)$ that are identical to 
those used in Fig.~14 of Ref.~\cite{thornburg}, and also $(p/M, e) = (2.250\,054\,2, 0.96)$ with $a/M = 0.995, 0.999\,9, 0.999\,95, 0.999\,99$, where the orbital parameters are $p$, the semilatus rectum and  $e$, the eccentricity. For the second step, we solve the inhomogeneous 
Teukolsky equation in the time-domain while feeding the trajectory information from the first step into 
the particle source-term of the equation. 

The numerical technique we use to solve the Teukolsky equation (\ref{teuk0}) is the same as the one 
presented in our earlier works (see Ref.~\cite{emri_code}). In particular, (i) we first rewrite the Teukolsky 
equation using compactified hyperboloidal coordinates that allow us to extract the evolving fields 
directly at null infinity  (${\mathscr{I}^+}$) while also solving the problem of artificial reflections from the outer boundary; 
(ii) we take advantage of axisymmetry of the background Kerr space-time, and separate the dependence 
on azimuthal coordinate, thus obtaining a set of (2+1) dimensional ($2+1$D) partial-differential-equations (PDEs); 
(iii) we then rewrite these equations into a first-order, hyperbolic PDE system; and in the last step (iv) 
we implement a two-step, second-order Lax-Wendroff, time-explicit, finite-difference numerical evolution 
scheme. The particle-source term on the right-hand-side of the Teukolsky equation requires some 
specialized techniques for such a finite-difference numerical implementation. Additional details can be 
found in our earlier works~\cite{emri_code} and the references included therein. The entire second step 
of the computation mentioned above is implemented using OpenCL/CUDA-based GPGPU-computing 
which allows for the possibility of performing very long duration and high-accuracy computations within 
a reasonable time-frame. Numerical errors in these computations are typically on the scale of a small 
fraction of a percent~\cite{xsede_paper}. 

Close to  the black hole our computational grid is approximately uniform in the $r^*$ radial coordinate. This is 
important for the high spin cases, because physically important quantities such as the innermost stable circular orbit (ISCO), the light ring, 
and the horizon tend to be very close to each other (in the ordinary $r$ coordinate), but are well-separated 
in $r^*$. For computations for sub-extremal Kerr spacetime it is typically sufficient to place the computational 
grid's inner boundary at a value  of ${r^*} \approx -100M$; but, this is not the case for the high spin case. In those cases the inner boundary must be located at a much larger negative value. In this current work, we use ${r^*} \approx -500M$. This, of course, makes the scale of the computation much larger and demanding.

\section{Numerical Results}


We first discuss the QNB in the $(a/M, p/M, e) = (0.999\,99, 2.918, 0.807)$ EMRI system, the same orbital parameters are used in Ref.~\cite{thornburg}. The motion is equatorial and prograde in all the cases we examine.  Throughout this Paper we choose the $(\ell,m)=(2,2)$ mode for a detailed analysis, where the $\ell,m$ mode is found by projecting out the $\ell$ mode from the sum over all $\ell$ modes that we find in our $2+1$D computation.

\begin{figure}
\includegraphics[width = 0.5\textwidth]{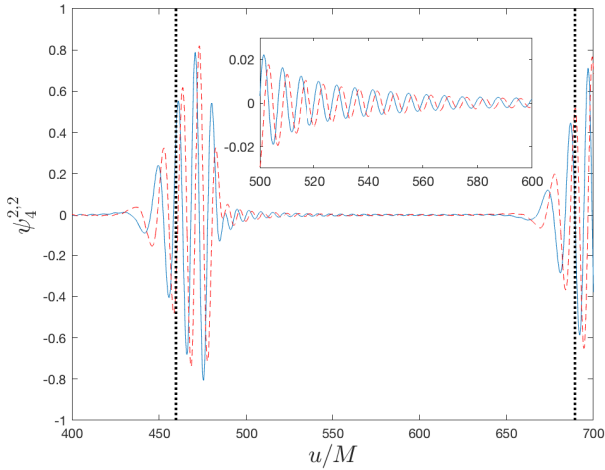}
\caption{The $\ell=2,m=2$ mode of the Weyl scalar $\psi_4$ as a function of retarded time $u$ at ${\mathscr{I}^+}$  from a high eccentricity EMRI system with a rapidly rotating central black hole. Here, the massive black hole has spin parameter $a/M=0.999\,99$, and the compact object's orbital parameters are $p/M=2.918$ and $e=0.807$. Solid (dashed) curve: $\Re(\psi_4^{2,2})$ ($\Im(\psi_4^{2,2})$) -- the real (imaginary) part of $\psi_4^{2,2}$. The vertical dotted lines show the time of periapsis passages. The inset zooms on the decaying wiggles. The frequency ratio here is $\rho\sim 34.5$. 
}
\label{wiggle}
\end{figure}

\begin{figure}[h]
\includegraphics[width = 0.5\textwidth]{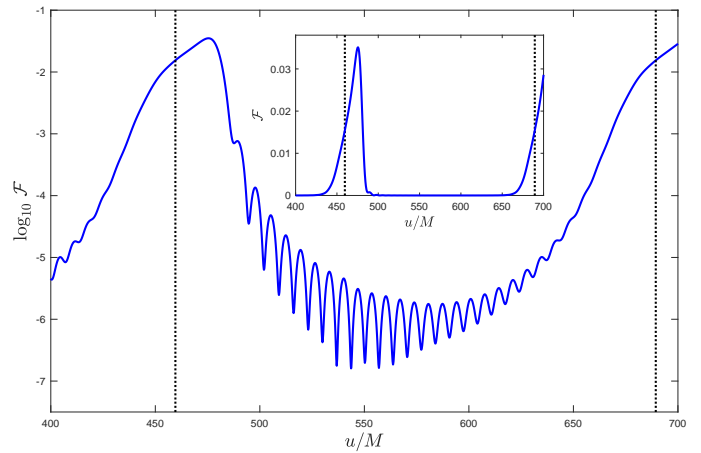}
\caption{The energy flux in gravitational waves $\cal{F}$ as measured  at ${\mathscr{I}^+}$ as a function of retarded time $u$ for the same parameters as in Fig.~\ref{wiggle}. The flux $\cal{F}$ is shown on a logarithmic scale to enhance the oscillations. The inset shows the same in a linear scale. The vertical dotted lines show the time of periapsis passages.  }
\label{en_flux}
\end{figure}

In Fig.~\ref{wiggle} we depict $\psi_4^{2,2}$ sampled at ${\mathscr{I}^+}$ as a function of retarded time $u$. We find good agreement with Fig.~14 in Ref.~\cite{thornburg} (note that there are some quantitative differences, likely due to different observing angles in use). We find that the oscillations in $\psi_4^{2,2} $  start shortly before periapsis passage, and are made of two main parts: First, high amplitude oscillations that are related to the whirl part of the orbit \cite{glampedakis:2002}, followed by the QNB (``wiggle") part, lingering high frequency oscillations that decay slowly until close to the next periapsis passage. Note that the QNB is high frequency, in the sense that the relevant time scale is much shorter than the orbital time scale between two consecutive periapsis passages. Defining the frequency ratio $\rho:=\omega/\Omega_{\rm orbital}$, we find for this case that $\rho\sim 34.1\gg1$. Therefore, the QNB frequencies are unrelated to the orbital frequency. Also, the QNB frequencies are separated from the whirl frequency of the zoom--whirl orbit which is seen close to the periapsis passage of the orbit. The same oscillations can also be seen in the flux of energy in gravitational waves $\cal{F}$   \cite{burko:2007}, which we show in Fig.~\ref{en_flux}. Notice that on a linear flux scale it is not easy to notice the QNB oscillations, and they are made clearer on a logarithmic scale. The average frequency that we measure from Fig.~\ref{en_flux} is $\omega_{\rm av}\sim 0.95M^{-1}$. While Fig.~\ref{wiggle} suggests that the periapsis passages are removed enough in time to study the QNB phenomenon in detail, Fig.~\ref{en_flux} shows that they are in fact too close to each other. Indeed, the QNB oscillations from one periapsis passage linger until after the QNB from the next periapsis passage starts, which makes the determination of the decay rate of the oscillations' amplitude difficult. 

The source of the QNB oscillations must therefore be different from either the orbital frequency or the whirl frequency, and the first candidate was proposed to be the QNM frequencies. That proposal was indeed confirmed for sub-extremal black holes \cite{nasipak}. (In Ref.~\cite{burko:2007} the orbital parameters were such that the orbital frequency and the QNM frequency were close, and therefore it was difficult to distinguish between them.) 
But a puzzle was found in the case that the central massive black hole is a nearly extreme Kerr black hole. Specifically, analysis of the instantaneous frequencies suggests that the frequency indeed varies from $0.93/M$ (near $u=500M$) to $0.98/M$ 
(near $u=600M$), as found in Ref.~\cite{thornburg}, where it is also noted that all the QNMs for this nearly-extreme Kerr black hole have frequencies $\sim0.995/M$ which is outside the frequency range relevant to the QNB waveform. This observation is the focus of this Paper.

\begin{figure}[h]
\includegraphics[width = 0.45\textwidth]{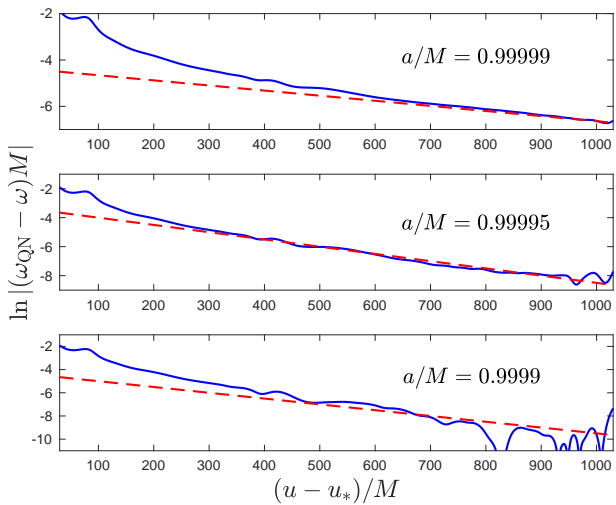}
\caption{QNB waveform frequency $\omega$ of $\psi_4^{2,2}$ depicted as functions of retarded time $u$ at ${\mathscr{I}^+}$ (solid curves), and predictions based on Eq.~(\ref{ang_freq}) (dashed lines).The EMRI system parameters used for the data presented here are $(p/M, e) = (2.250\,054\,2, 0.96)$ with $a/M = 0.999\,9, 0.999\,95, 0.999\,99$.}
\label{wiggle_w}
\end{figure}

To study this apparent discrepancy in the frequencies in greater detail we specialize next to orbital parameters that result in longer duration and higher amplitude QNB. Specifically, we choose $(p/M, e) = (2.250\,054\,2, 0.96)$ with $a/M = 0.995, 0.999\,9, 0.999\,95, 0.999\,99$. 
Figure \ref{wiggle_w} depicts $\omega_{QN}-\omega$ for several $a/M$ values extracted from the waveform at ${\mathscr{I}^+}$ as functions of  retarded time $u$. Here, $\omega$ is the instantaneous  frequency of the QNB and $\omega_{\rm QN}$ is the [real part of the $n=0$ (``fundamental'') mode of the] QNM frequency, computed with the code of Ref.~\cite{stein}. (The various $n$ overtones are all grouped very near to each other and specifically to the $n=0$ mode.)  From Ref.~\cite{andy} we find   that for plunge trajectories the (real part of the) angular frequency for each $\ell,m$ mode, after summing over all overtones, is given by 
\begin{equation}\label{ang_freq}
\omega_{\ell,m}(u)=\frac{m}{2M} - \frac{3m}{8M} \times e^{-\kappa(u-u_0)}\, ,
\end{equation}
where $\kappa$ is the black hole's surface gravity. 
Motivated by this result, we propose that when summing over all overtones for the eccentric orbits we consider here, $\omega = \omega_{QN} - \alpha \exp(-\beta u)$ where we predict the parameter $\beta\approx \kappa$. Notice, that for nearly-extreme black holes the surface gravity $\kappa= \frac{1}{2M}\,\sqrt{1-(a/M)^2}+O[1-(a/M)^2]$, so that $\beta\approx \frac{1}{2M}\,\sqrt{1-(a/M)^2}$, with which we find in Fig.~\ref{wiggle_w} good numerical agreement. 

Figure \ref{wiggle_w} suggests that the QNB frequency is a monotonically increasing function of time, that is likely to approach the QNM frequency at very late times. Notice, however, that 
because of the periodic motion of the compact object the next periapsis passage interferes with the relaxation of the QNB frequency, and resets it. To see the agreement of the late-time QNB frequency with the massive black hole's QNM frequency better one would need to have a longer lapse of time between two consecutive periapsis passages. However, it is harder to model the latter numerically as it requires higher eccentricity values, and consequently cause the apoapsis to be farther out. The latter point would then require moving the hyperboloidal layer of the computational domain farther out, resulting in significantly increased computational time.

\begin{figure}
\includegraphics[width = 0.5\textwidth]{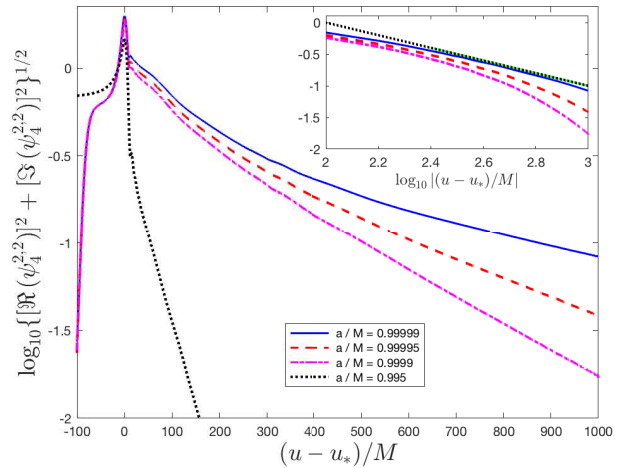}
\caption{QNB waveform amplitude of $\psi_4^{2,2}$ plotted at ${\mathscr{I}^+}$ as a function of retarded time $u$ for the same orbital parameters as in Fig.~\ref{wiggle_w}, for $a/M = 0.995, 0.999\,9, 0.999\,95, 0.999\,99$. The inset shows the same in a log-log plot, with an added straight reference line whose slope is $-1$. The case $a/M=0.995$ is excluded from the variable range shown in the inset. The parameter $u_*$ is chosen such that the peak of the amplitude is at $u-u_*=0$. } 
\label{wiggle_amp}
\end{figure}

Figure \ref{wiggle_amp} shows the amplitudes of the QNB signals for a number of $a/M$ values for the same orbital parameters as in Fig.~\ref{wiggle_w} as functions of retarded time $u$. We infer that the QNB amplitude has a transient power-law decay rate, that is consistent with the expected decay rate $\sim u^{-1}$. In addition to the next periapsis passage that makes the available time series for determination of the decay rate finite and short, we have an additional complication from the transient nature of nearly-extreme black holes. Specifically, even close to extremality eventually the nearly-extreme black hole behaves like sub-extreme ones \cite{nek2,burko:2019}. The decay rate of the QNB amplitude deviates as a transient behavior from the exponential decay rate of a single QNM because of the many overtones that contribute to it. But with increasing times one mode ($n=0$) become relatively more dominant, such that eventually the transient behavior would transition to an asymptotic behavior of exponential decay of a single QNM if the next QNB were not to interfere. 
The latter effect explains the curving of the amplitude from the straight line in the inset of Fig.~\ref{wiggle_amp} at late times. We emphasize that while the inverse (retarded)  time behavior of the amplitude is by no means a new phenomenon, and is closely related to the behavior found in Refs.~\cite{yang:2013,glampedakis:2001,nek2}, the time dependence of the frequency is new, to the knowledge of the present authors. 

\begin{figure}
\includegraphics[width = 0.45\textwidth]{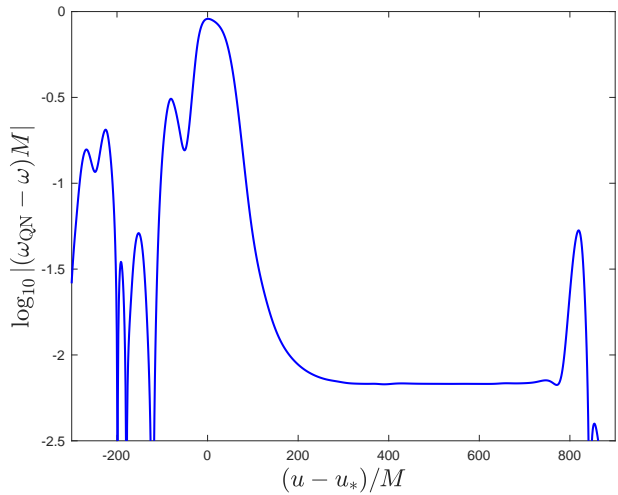}
\caption{Waveform frequency $\omega$ of $\psi_4^{2,2}$ depicted at ${\mathscr{I}^+}$ as a function of retarded time $u$ . The parameter $\omega_{\rm QN}$ is the QNM frequency of the $n=0$ overtone calculated with Ref.~\cite{stein}. The EMRI system parameters used for the data presented here are $(p/M, e) = (2.250\,054\,2, 0.96)$ with $a/M = 0.995$. } 
\label{subextreme}
\end{figure}

Figure \ref{subextreme} shows the frequency of the radiation for the same orbital parameters as above for the case $a/M=0.995$. Notice that the variability in the frequency of the high frequency QNB is not present. Instead, we see the frequency approach that of a single QNM, to 6 parts in $10^3$, the accuracy of this computation. 

We infer that for nearly-extreme Kerr black holes the QNB part of the waveform is not likely to be representable by a single QNM: (i) the frequency is variable, and (ii) the amplitude does not decay exponentially with 
time. Instead, the QNB waveform is likely a suitable mixture of many QNMs, such that higher overtones decay faster with time and the fundamental $n=0$ mode gradually dominates more than higher overtones. 

\begin{figure}
\includegraphics[width = 0.45\textwidth]{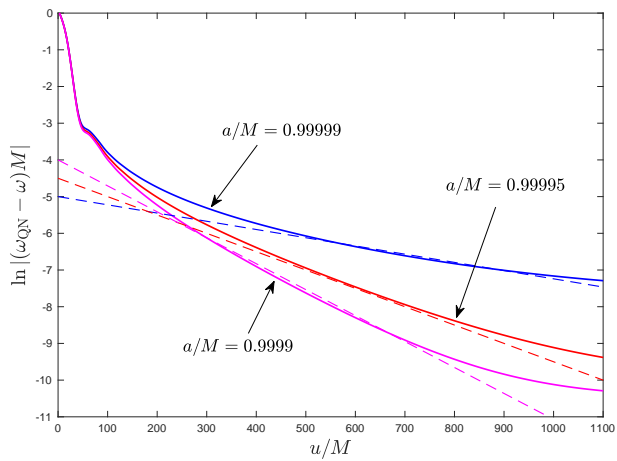}
\caption{The waveform frequency $\omega_{QN}-\omega$ plotted at ${\mathscr{I}^+}$ as a function of retarded time $u$  for a source free scalar field, for $a/M = 0.999\,9, 0.999\,95, 0.999\,99$ (solid curves). The dashed straight lines are reference lines corresponding to the frequencies predicted from  Ref.~\cite{andy}.}
\label{scalar_w}
\end{figure}

\begin{figure}
\includegraphics[width = 0.45\textwidth]{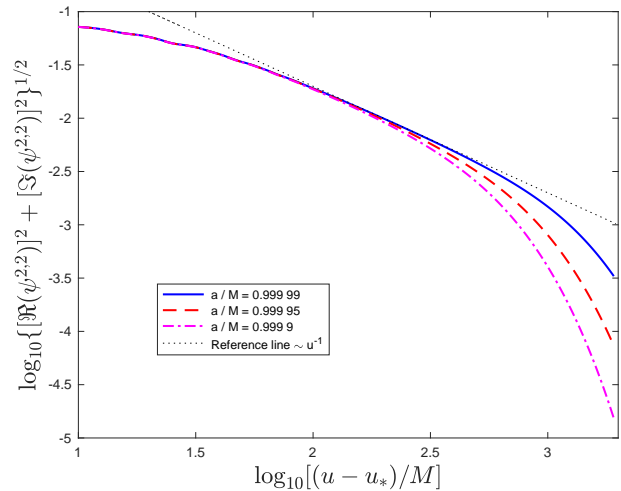}
\caption{The waveform amplitude plotted as a function of retarded time $u$ at ${\mathscr{I}^+}$ for the $\ell,m=2,2$ mode of a source free scalar field $\psi$ for the following $a/M$ values: $a/M = 0.999\,9$ (dash-dotted curve), $0.999\,95$ (dashed curve), and $0.999\,99$ (solid curve). The dotted straight line is a reference line with slope $\sim u^{-1}$. The parameter $u_*$ is chosen so that the maximum of the amplitude is at $u-u_*=0$.}
\label{scalar_amp}
\end{figure}

Next, we bring forth evidence for the origin of the QNB phenomenon. Specifically, we consider a source-free field propagating on the fixed background of a nearly-extreme Kerr black hole. Consider for simplicity the freely propagating $\ell,m=2,2$ mode of a scalar field $\psi$ ($s=0$). We set the initial data to be an initially stationary Gaussian wave packet centered at $r/M = 1$ of width $0.1M$. Figure \ref{scalar_w} shows the scalar wave frequency $\omega$ as a function of retarded time $u$. We note that the transient phenomenon in Fig.~\ref{scalar_w} is very similar to that seen in Fig.~\ref{wiggle_w}, and therefore we suggest that the presence of the compact object is not significant for the generation of the QNB: instead, the crucial factor that generates the QNB is the strong excitation of QNM, whether it is done by a particle or by free fields. Indeed, the generation of QNB is insensitive to the details of the particle's motion when there is a particle moving: In Refs.~\cite{andy,geoff} the particle plunges into the nearly-extreme black hole, instead of being in an orbit, and still the angular frequency evolution is similar, which allows us to compare our angular frequencies with the prediction based on Ref.~\cite{andy,geoff}. We display the amplitude of $\psi$ in Fig.~\ref{scalar_amp} as a function of retarded time $u$ for a number of $a/M$ values. The behavior of the free scalar field amplitude is very similar to the behavior of the amplitude of the gravitational perturbations for the orbital motion of a particle course (cf.~Fig.~\ref{wiggle_amp}). This conclusion is not unexpected, as the analysis in Ref.~\cite{nek2} suggests that the transient decay rate for such a field would be $\sim u^{-1}$ because of the excitation of many quasi-normal overtones. The asymptotic faster decay rate at late retarded times is due to the nearly-extreme nature of the central black hole: the asymptotic behavior is that of a sub-extreme black hole. 

\section{Concluding remarks}

There are some differences between the two QNB sources that we consider: When the source is a compact object whose orbital periapsis gets close enough to the black hole to strongly excite many QNM overtones that results in QNB, the repeated periapsis passages prevent the late time behavior from appearing. However, when the QNM excitation source is a free field, the latter first excites many QNM overtones and therefore at first looks similar to the QNB, but later the field is absorbed by the black hole or disperses to ${\mathscr{I}^+}$ and therefore does not excite the overtones anymore, and therefore eventually decays at the same rate as that of the least damped overtone. Notice, that as only a single QNM dominates over the field at late times we can determine its frequency only to a certain accuracy (cf. Fig.~\ref{subextreme}). Indeed, in Fig.~\ref{scalar_w} at late times the frequencies curve up, consistently with this picture. No similar curving is seen in Fig.~\ref{wiggle_w}, because the next QNB overwhelms the signal.

We find the signal to start before periapsis passage: As the compact object gets close to the periapsis it already moves in the strong-field region in the whirl part of the orbit, where gravitational waves are excited at high amplitudes. Therefore, we expect the retarded time value of the signal to start slightly before periapsis passage, assuming the prompt (unscattered) signal moves along geometrical optics rays. The whirl part of the signal is then followed by the lingering QNB signal, that persists through the zoom part of the orbit until the next whirl signal because of the slow (inverse time) decay of its amplitude.

We note that the peculiar aspects of the gravitational waveform in nearly-extreme Kerr black holes ~\cite{yang:2013,nek2} -- e.g., the QNM portion of the scalar or gravitational waveform amplitude
decays as $M/u$ for near extremal Kerr black holes due to the unusual ``stacking'' of QNM overtones -- is closely related to the QNB waveform. Moreover, the transient behavior of radiation fields in nearly-extreme Kerr black holes \cite{burko:2019} is closely related to the transient behavior found herein. 

The expression of the time-dependence of the frequencies of the gravitational waveform in the context of a small body plunging into a near extremal Kerr black hole as a sum over QNM overtones~\cite{andy} provides further evidence that the explanation for the unexpected features present in the QNB waveform lies in the sum over the QNM overtones of a near-extremal Kerr black hole.



In contrast with sub-extremal black holes, for which the QNM portion of the signal for a specific $(\ell,m)$ mode has a fixed frequency (cf.~Fig.~\ref{subextreme}), in the near-extremal case the QNB frequency has a rich time dependence arising from the excitation of overtones. These excitations  suggest that in addition to the QNM mode amplitudes, there is additional information about the binary system that may be extracted from the details of the time dependence of the frequency. If near-extreme  black holes exist, and are discovered by current or  future gravitational wave observatories (see, e.g., Ref.~\cite{ligo-nek}), our work suggests that one may have to perform a very detailed template signal study in order to extract the type of information that may be coded in the QNM portion of the signal. The repeated excitation of the QNMs with each periapsis passage, in addition to the long-lived transient waveform (which decays as inverse time), bears the potential for highly accurate observational determination  of such information. 

\section*{Acknowledgments}


We thank Chuck Evans, Zach Nasipak, and Jonathan Thornburg for making valuable comments on an earlier draft of this Paper. 
G.~K.~acknowledges research support from National Science Foundation (NSF) Grants Nos. PHY-1701284 and DMS-1912716, and Office of Naval Research/Defense University Research Instrumentation Program (ONR/ DURIP) Grant No.~N00014181255.


\begin{thebibliography}{11}

\bibitem{thornburg}
J.~Thornburg, B.~Wardell, and M.~van de Meent, arXiv:1906.06791 [gr-qc].

\bibitem{nasipak} Z.~Nasipak, T.~Osburn, and C.R.~Evans, Phys.~Rev.~D {\bf 100}, 064008 (2019).

\bibitem{burko:2007} L.M.~Burko and G.~Khanna, Europhys.~Lett.~{\bf 78}, 60005 (2007). 

\bibitem{hughes:2016} S.~O'Sullivan and S.A.~Hughes, Phys.~Rev.~D {\bf 94}, 044057 (2016). 

\bibitem{thornburg:2017} J.~Thornburg and B.~Wardell, Phys.~Rev.~D {\bf 95}, 084043 (2017). 



\bibitem{kojima}
Y.~Kojima and T.~Nakamura, Prog.~Theor.~Phys.~{\bf 72}, 494 (1984).

\bibitem{emri_code}
P.A.~Sundararajan, G.~Khanna, and S.A.~Hughes, Phys.~Rev.~D {\bf 76}, 104005 (2007); P.A.~Sundararajan, G.~Khanna, S.A.~Hughes, and
S.~Drasco, Phys.~Rev.~D {\bf 78}, 024022 (2008); P.A. Sundararajan, G. Khanna, and S.A. Hughes, Phys. Rev. D {\bf 81}, 104009 (2010); A.~Zengino\u{g}lu and G.~Khanna, Phys.~Rev.~X {\bf 1}, 021017 (2011).

\bibitem{xsede_paper}
J. McKennon, G. Forrester, and G. Khanna, Proceedings
of the NSF XSEDE12 Conference, Chicago (2012).

\bibitem{stein} L.~Stein, arXiv:1908.10377 [astro-ph.IM].

\bibitem{glampedakis:2002} K.~Glampedakis and D.~Kennefick, Phys.~Rev.~D {\bf 66} 044002 (2002). 

\bibitem{andy}
S.~Hadar, A.P.~Porfyriadis, A.~Strominger, Phys.~Rev.~D {\bf 90}, 064045 (2014).

\bibitem{burko:2019} L.M.~Burko, G.~Khanna,~and S.~Sabharwal, arXiv:1906.03116 [gr-qc].

\bibitem{nek2}
L.M.~Burko and G.~Khanna, Phys.~Rev.~D {\bf 94}, 084049 (2016).

\bibitem{yang:2013} H.~Yang, A.~Zimmerman, A.~Zengino\u{g}lu, F.~Zhang, E.~Berti, and Y.~Chen, Phys.~Rev.~D {\bf 88}, 044047 (2013).

\bibitem{glampedakis:2001} K.~Glampedakis and N.~Andersson, Phys.~Rev.~D {\bf 64},
104021 (2001).







\bibitem{geoff}
G.~Compere, K.~Fransen, T.~Hertog, and J.~Long, Class.~Quantum Grav.~{\bf 35}, 104002 (2018).

\bibitem{ligo-nek}
B.~Zackay, T.~Venumadhav, L.~Dai, J.~Roulet, and M.~Zaldarriaga, Phys. Rev. D {\bf 100}, 023007 (2019).




\end{thebibliography}
\end{document}